\documentclass[doublespacing]{elsart}

\usepackage{graphicx}
\usepackage{amssymb}

\begin{document}

\begin{frontmatter}

\title{Substrate effects on V$_2$O$_3$ thin films}

\author{Udo Schwingenschl\"ogl$^{1,2,3}$\corauthref{cor1}},
\corauth[cor1]{Corresponding author. E-Mail: schwinud@physik.uni-augsburg.de}
\author{Raymond Fr\'esard$^3$}, and
\author{Volker Eyert$^{3,4}$}
\address{$^1$ICCMP, Universidade de Bras\'ilia, 70904-970 Bras\'ilia, DF, 
         Brazil}
\address{$^2$Theoretische Physik II, Institut f\"ur Physik, 
         Universit\"at Augsburg, 86135 Augsburg, Germany}
\address{$^3$Laboratoire CRISMAT, UMR CNRS-ENSICAEN(ISMRA) 6508, and 
         IRMA, FR3095, 14050 Caen, France}
\address{$^4$Center for Electronic Correlations and Magnetism, 
         Institut f\"ur Physik, 
         Universit\"at Augsburg, 86135 Augsburg, Germany}

\begin{abstract}
We apply density functional theory and the augmented spherical wave 
method to analyze the electronic structure of V$_2$O$_3$ in the 
vicinity of an interface to Al$_2$O$_3$. The interface is modeled by a 
heterostructure setup of alternating vanadate and aluminate slabs. We 
focus on the possible modifications of the V$_2$O$_3$ electronic states in 
this geometry, induced by the presence of the aluminate layers. 
In particular, we find that the tendency of the V $ 3d $ states to 
localize is enhanced and may even cause a metal-insulator transition. 
\end{abstract}

\begin{keyword}
density functional theory \sep electronic structure \sep 
V$_2$O$_3$/Al$_2$O$_3$ heterostructure
\PACS 73.20.-r \sep 73.21.Ac \sep 73.61.-r
\end{keyword}
\end{frontmatter}

\section{Introduction}

V$_2$O$_3$ is a classical example of a material which is located in the 
vicinity of a Mott transition. The latter can be triggered, for example, 
by Cr-substitution \cite{kuwamoto80} or by the application of external 
pressure \cite{limelette03}. In this respect depositing thin films of 
V$_2$O$_3$ on a substrate with a slightly smaller in-plane lattice 
parameter $ a $ is likely to provide another way of inducing a Mott 
insulator-to-metal transition. Indeed, V$_2$O$_3$ thin films are 
nowadays attracting great attention, since they promise further insight 
into the mechanism of the Mott transition \cite{hague08,allimi08}. In this context, 
the observation of a thickness-dependent metal-insulator transition (MIT) 
in ultrathin films has been attributed to an increasing $c$/$a$ lattice 
parameter ratio (decreasing $a$, increasing $c$) because of
interaction with the Al$_2$O$_3$ substrate \cite{luo04}. In contrast, detailed 
X-ray diffraction experiments for high quality films indicate a more 
complex thickness-dependence of the lattice parameters \cite{grygiel07}. 
In particular, they do not point to systematic changes of the $ c $ 
lattice constant for film thicknesses ranging from 100\,\AA\ to 1000\,\AA.

In this Letter, we aim at clarifying the influence of the substrate 
on the electronic states of V$_2$O$_3$ thin films grown on Al$_2$O$_3$. 
Indeed, for ultrathin films a significant reorganization of the 
electronic excitations of the film material may be induced by those 
of the substrate, going beyond geometric effects. In general, the
properties of heterointerfaces in many cases deviate considerably from
those of the component materials. A prominent example in this context
is the quasi two-dimensional electron gas formed at the interface
of the insulators LaAlO$_3$ and SrTiO$_3$ \cite{ohtomo04,nakagawa06,gemming06}.
However, as we will show below, the impact of an Al$_2$O$_3$ interface 
on the near-$ E_F $ density of states in V$_2$O$_3$ is rather weak as 
soon as the heterostructure involves several V$_2$O$_3$ layers.

To describe the electronic effects, we model the interface as a 
heterostructure consisting of alternating slabs of the component 
materials. Comparing the V $3d$ partial densities of states as 
resulting from the heterostructure calculation to the corresponding 
bulk data we find only a weak response of the vanadate to the bonding 
with the aluminate. However, the reduced dispersion of the V $ 3d $ 
states perpendicular to the layers leads to a narrowing especially 
of the $ a_{1g} $ bands, which might suppress the intrinsic metalicity 
of $ {\rm V_2O_3} $. 

\section{Structural and calculational details}

The unit cell used for modeling the heterostructure is the canonical 
threefold hexagonal supercell of the trigonal corundum lattice. It 
comprises six formula units as well as a stacking of six metal layers 
along the hexagonal $c$ axis. These layers are separated by layers
of O atoms, which form octahedra around the metal atoms. In order to 
set up the heterostructure, we divide the hexagonal supercell into a 
V$_2$O$_3$ domain and an Al$_2$O$_3$ domain, both consisting of three metal 
layers as well as the corresponding O layers. The $c$ lattice constants 
and the fractional coordinates of the atoms in the two slabs are those of 
bulk metallic V$_2$O$_3$  \cite{dernier70} and Al$_2$O$_3$ \cite{toebbens01}, 
respectively. In contrast, since V$_2$O$_3$ is grown on an Al$_2$O$_3$ 
substrate, we adopt the $a$ lattice constant of the aluminate for the 
whole supercell. Because the $a$ lattice constant of bulk V$_2$O$_3$ 
is about 3.5\% larger as compared to that of bulk Al$_2$O$_3$, a 
relevant interface stress is induced, which could modify the V valence 
states. However, no structural relaxation has been found in the 
experiments \cite{grygiel07}. Since a minor relaxation beyond
the experimental resolution does not affect our further conclusions, a
structure optimization consequently is dispensable.

\begin{figure}[b]
\centering 
\includegraphics*[width=0.3\textwidth,clip]{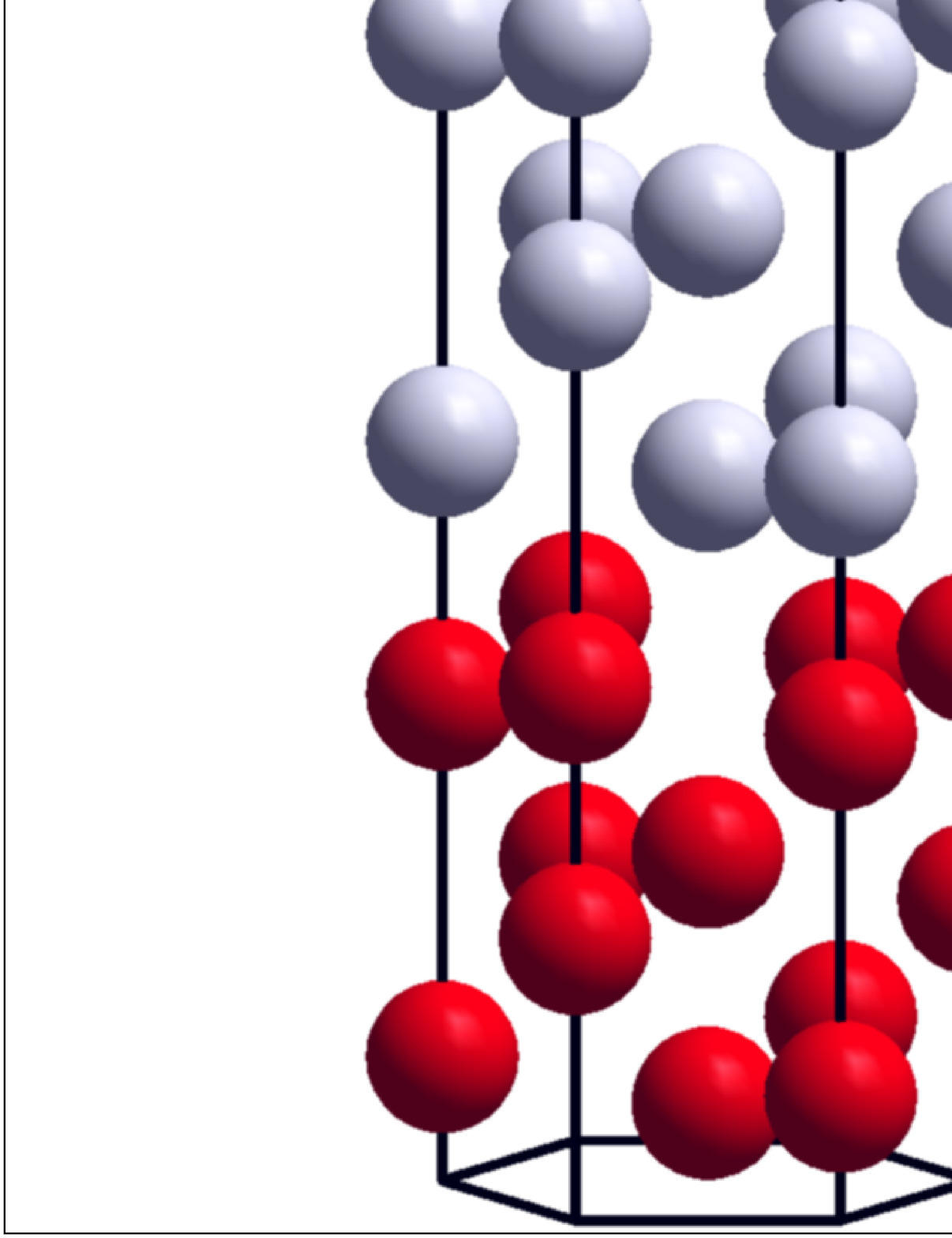}
\caption{Hexagonal unit cell of the V$_2$O$_3$/Al$_2$O$_3$
         hetero\-structure, as used in the calculations. The light gray
         and red (b/w: dark gray) spheres denote V and Al sites,
         respectively. The figure shows six buckled (horizontal) planes
         of V/Al atoms. For clarity, the O atoms have been omitted. 
         They form the well-known O sublattice of corundum V$_2$O$_3$
         and Al$_2$O$_3$, see the detailed description in Ref.\ 
         \cite{dernier70}.}
\label{fig1}
\end{figure}

Finally, using periodic boundary conditions we obtain a multilayer 
structure of alternating V$_2$O$_3$ and Al$_2$O$_3$ slabs, as
illustrated in Fig.\ \ref{fig1}. 
The O atoms, which form a sublattice 
of space-filling octahedra, are omitted in this representation, since 
they maintain the positions of the parent compounds. For further 
details of the structural aspects of V$_2$O$_3$ and related oxides 
see Ref.\ \cite{oktaeder} as well as the references given therein.
We have checked that our results do not depend on the supercell
setup, i.e.\ the thickness of the supercell in $c$ direction and the
thickness of the V$_2$O$_3$ slab.

Our investigation is based on density functional theory within the 
local density approximation. We use a new full-potential version 
of the augmented spherical wave method \cite{aswrev,aswbook}. The 
basis set contains Al $3s$, $3p$, $3d$, O $2s$, $2p$, and V $4s$, 
$4p$, $3d$ orbitals and is complemented by states of additional 
augmentation spheres. Brillouin zone integrations were performed 
using the linear tetrahedron method on a mesh of up to 264 {\bf k}-points 
within the irreducible wedge of the hexagonal zone. 

We point out, that the well-known shortcomings of the local density 
approximation necessitate a careful analysis of the electronic 
structure in order to derive reliable conclusions as strong 
correlation effects are largely ignored in this approach. As has 
been pointed out by Brinkman and Rice and elaborated by others 
\cite{brinkman,spalek,weber,lamboley}, such correlation effects may 
affect the metallic behavior especially if narrow bands are present 
near the Fermi energy. Thus, the following line of reasoning 
is closely related to previous work on the phase transitions
in vanadium and titanium oxides \cite{us0304,cpl04,eyert05} and 
influenced by the results of recent LDA+DMFT calculations for bulk 
$ {\rm V_2O_3} $ \cite{held01,keller04}.

\section{Results and discussion}

\begin{figure}[b]
\centering 
\includegraphics*[width=0.5\textwidth,clip]{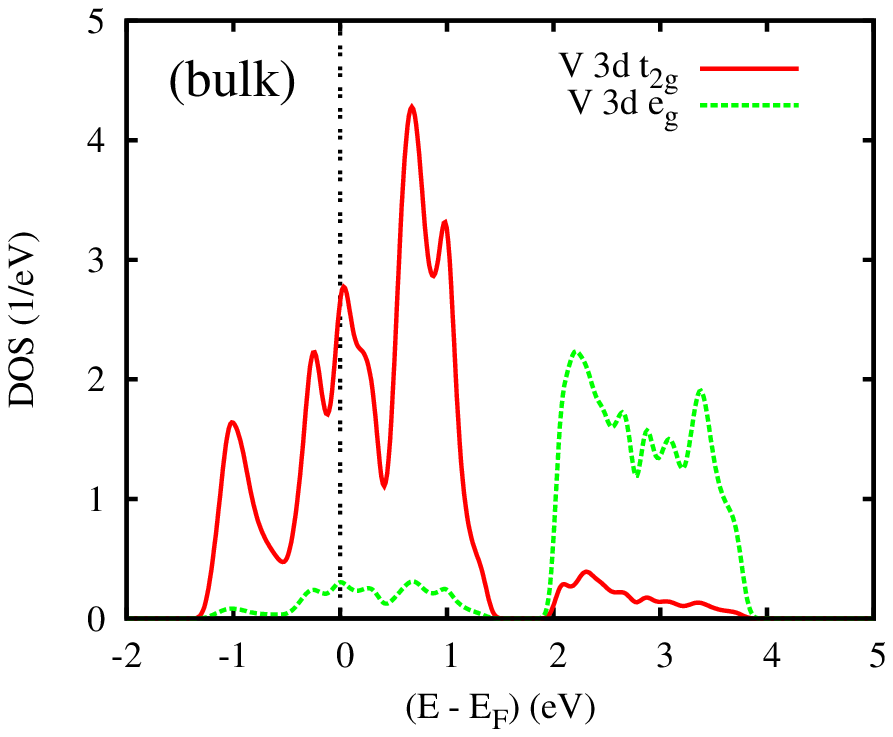}\\[0.2cm]
\includegraphics*[width=0.5\textwidth,clip]{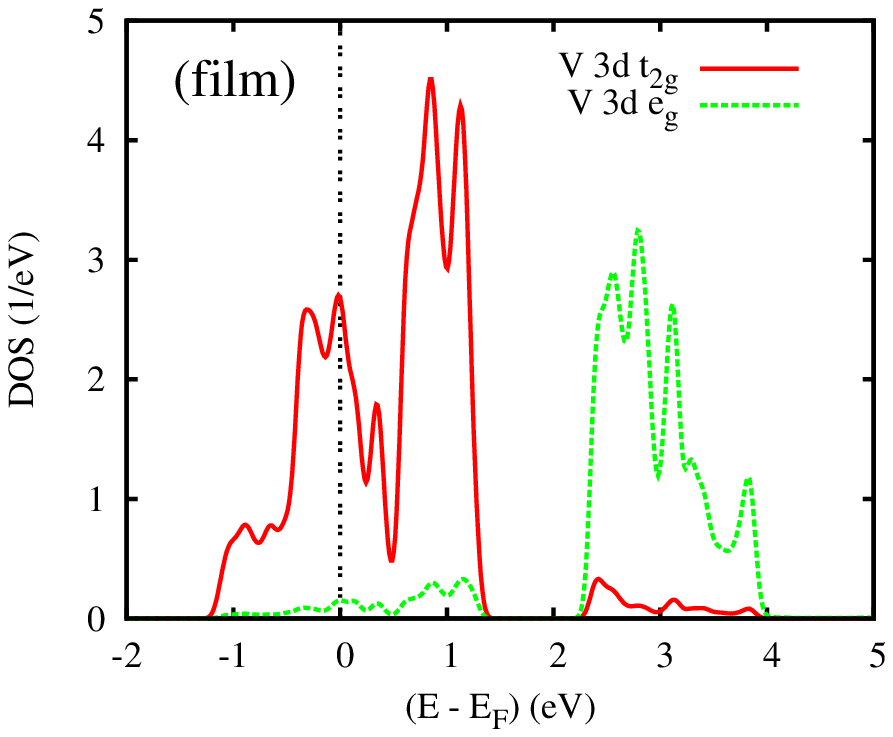}
\caption{Partial V $3d$ DOS per atom for the metallic phase of V$_2$O$_3$ 
(upper panel) and for a V$_2$O$_3$/Al$_2$O$_3$ heterostructure (lower panel).
The lower panel, representing the average of all V atoms, indicates a band
narrowing which affects the whole V$_2$O$_3$ slab. Besides, the data point to
only minor alterations of the electronic states in a V$_2$O$_3$ thin film because
of the contact to an Al$_2$O$_3$ substrate.}
\label{fig2}
\end{figure}

Turning to the results of our calculations, we compare in Fig.\ \ref{fig2}
the V $3d$ partial densities of states (DOS) of metallic bulk V$_2$O$_3$ to 
those of the heterostructure. For a more detailed discussion of the band 
structure and DOS of V$_2$O$_3$ \cite{eyert05} and related vanadates 
\cite{oktaeder} we refer the reader to the literature. According to 
Fig.\ \ref{fig2}, even for a V$_2$O$_3$ triple-layer the essential 
features of the V $3d$ electronic states largely resemble those of 
bulk V$_2$O$_3$. Besides minor differences of the DOS shape, the 
energetic positions of the main peaks are very similar. This holds for 
both the (partially occupied) $t_{2g}$ and the (unoccupied) $e_g$ 
group of states. Importantly, these findings are rather independent 
of small variations of the V positions, which reflects the strong 
localization of the V $3d$ states \cite{poteryaev07}. In addition, 
the similarity of the bulk and layer partial densities of states 
builds on the fact that the $p$-$d$ hybridization resulting from 
the covalent bonding between V $3d$ and O $2p$ orbitals is similar 
for both geometries. 

In passing, we mention that the Al electronic states resemble those 
of the bulk aluminate. This is due to the fact that Al$_2$O$_3$ is an 
insulator with a large band gap of about 8.7\,eV. For the
heterostructure the Al states thus are found far below and above the Fermi
energy, and are hardly influenced by the V states. 

The most important difference between the bulk and triple-layer partial 
densities of states arises from the slightly reduced bandwidths of the 
latter, again for both the $t_{2g}$ and $e_g$ bands. A similar band 
narrowing has been observed for the $ t_{2g} $ bands of Cr-doped bulk 
V$_2$O$_3$, where it traces back to a reduction of the $d$-$d$ overlap 
coming along with the antidimerization of the V--V pairs parallel to 
the hexagonal $c$-axis. In that case, LDA+DMFT calculations have clearly 
demonstrated that the band narrowing, as induced by the structural changes,
drives the metal-insulator transition as soon as the strong electronic 
correlations, which are only rudimentary covered by the LDA, are 
properly included \cite{held01,keller04}. Transferring that result to 
the present situation, we may likewise expect insulating behavior 
from the reduced bandwidth of triple-layer $ {\rm V_2O_3} $. 

This finding should be related to the fact that the $ a $ lattice constant 
used here is by 3.5\% smaller than that of bulk metallic $ {\rm V_2O_3} $ 
and, hence, would be rather in favor of metallic behavior. In addition, 
the $ c $ lattice constant and the fractional coordinates of the atoms in 
the slab are identical to those of bulk metallic $ {\rm V_2O_3} $, which 
again would preferably lead to metallic behavior. As a consequence, from 
purely geometric considerations we would expect a broadening of the 
$ t_{2g} $ bands as compared to bulk metallic $ {\rm V_2O_3} $, which 
would be in favor of metallicity even in the presence of strong 
electronic correlations. 

Yet, the opposite situation occurs. From this we conclude that the band 
narrowing in the triple-layer is mainly caused by the reduced $d$-$d$ 
overlap due to the restricted layer-geometry and only to a lesser degree 
by lattice parameters and atomic parameters. In fact, variation of the
$a$ lattice constant (within a reasonable range) has virtually no effect
on the magnitude of the band narrowing. As a consequence, taking 
into account the effect of strong electronic correlations we would expect 
insulating behavior for rather thin films and a tendency towards metallicity 
for thicker films. Indeed, this trend has been observed in the 
experiments \cite{allimi08,luo04,grygiel07,grygiel08}.

\section{Conclusion}

In conclusion, electronic structure calculations for a 
V$_2$O$_3$/Al$_2$O$_3$ heterostructure point to rather localized 
V $3d$ $t_{2g}$ states as compared to the situation in bulk 
metallic $ {\rm V_2O_3} $. Although the interface hardly affects the 
V$_2$O$_3$ electronic states, reduction of the $t_{2g}$ band width 
due to reduced $d$--$d$ overlap perpendicular to the layers may 
cause an insulating ground state of the system if the strong 
electronic correlations are taken into account. While this mechanism 
applies mainly to thin films with a thickness of a few V$_2$O$_3$ unit 
cells, increase of the film thickness will more likely drive an 
insulator-to-metal transition. 

\section*{Acknowledgments}
We acknowledge discussions with C.\ Grygiel and T.\ Kopp, as well as 
financial support by the Deutsche Forschungsgemeinschaft through 
SFB 484 and by the Bayerisch-Franz\"osisches Hochschulzentrum. 
Fig.\ \ref{fig1} was generated using the XCrysDen software package by 
A.\ Kokalj [Comp.\ Mater.\ Sci.\ {\bf 28}, 155 (2003)].

\end{document}